\documentclass[showpacs,preprintnumbers,amsmath,amssymb]{revtex4}
\usepackage[dvips]{epsfig}
\usepackage{latexsym}
\usepackage{psfrag}
\def\setb@se#1{\baselineskip=#1 \normalbaselineskip=#1}

\newcommand{\be}{\begin{equation}}
\newcommand{\ee}{\end{equation}}
\newcommand{\beqn}{\begin{eqnarray}}
\newcommand{\eeqn}{\end{eqnarray}}
\newcommand{\bsub}{\begin{subeqnarray}}
\newcommand{\esub}{\end{subeqnarray}}

\newcommand{\bea}{\begin{eqnarray}}
\newcommand{\ea}{\end{eqnarray}}
 
\newcommand{\ba} {\begin{eqnarray}}
\newcommand{\e} \epsilon
\newcommand{\la} \lambda
\newcommand{\La} \Lambda

\newcounter{subequation}[equation]

\makeatletter \expandafter\let\expandafter\reset@font\csname
reset@font\endcsname
\newenvironment{subeqnarray}
  {\arraycolsep1pt
    \def\@eqnnum\stepcounter##1{\stepcounter{subequation}{\reset@font\rm
      (\theequation\alph{subequation})}}\eqnarray}%
  {\endeqnarray\stepcounter{equation}}
\makeatother

\begin{document}

\title{An analytical approximation scheme to two point boundary value problems
of ordinary differential equations}
\author{Bruno Boisseau$^{1}$, P\'eter Forg\'acs$^{1,2}$ and Hector Giacomini$^{1}$}
 \affiliation{ {$^{1}$Laboratoire de Math\'{e}matiques et Physique Th\'{e}orique CNRS\\
 \hbox{Universit\'{e} Fran\c ois Rabelais Tours, F\'ed\'eration Denis Poisson}\\
Parc de Grandmont, 37200 Tours, France}\\
 {$^{2}$MTA RMKI, H-1525 Budapest, P.O.Box 49, Hungary}
}

 \begin{abstract}
A new (algebraic) approximation scheme to find {\sl global} solutions of
two point boundary value problems of
ordinary differential equations (ODE's) is presented.
The method is applicable for both linear and
nonlinear (coupled) ODE's whose solutions are analytic near one of the boundary points.
It is based on replacing the original ODE's by a sequence of auxiliary first order
polynomial ODE's with constant coefficients. The coefficients in the auxiliary ODE's are
uniquely determined from
the local behaviour of the solution in the neighbourhood of one of the boundary points.
To obtain the parameters of the global (connecting) solutions analytic at one of the
boundary points, reduces to find the appropriate zeros of algebraic equations.
The power of the method is illustrated by computing the approximate values of the ``connecting
parameters'' for a number of nonlinear ODE's arising in various problems in field theory.
We treat in particular the static and rotationally symmetric global vortex, the skyrmion,
the Nielsen-Olesen vortex, as well as the 't Hooft-Polyakov magnetic monopole.
The total energy of the skyrmion and of the monopole is also computed by the new method.
We also consider some ODE's coming from the exact renormalization group.
The ground state energy level of the anharmonic oscillator is also computed
for arbitrary coupling strengths with good precision.

\end{abstract}
\maketitle
It occurs quite often in Physics (but of course also in other
areas of Science) that we have to solve (singular) two point
boundary value problems (2p.\ BVP) associated to a system of
ODE's. An important class of such BVP's for linear ODE's arises
from eigenvalue problems of the stationary Schr\"odinger equation
either in one dimension or reduced to an ODE with some (e.g.\
rotational) symmetry. Another large class of 2p.\ BVP's for {\sl
nonlinear} ODE's stems from the equations of motion of classical
field theories reduced to ODE's (e.g.\ with some symmetries) and
one could easily continue the list. We start with the example of the
static, rotationally symmetric global vortex in a Ginzburg-Landau effective
theory \cite{ginsb-pit58}, which has numerous applications ranging from condensed matter
to cosmic strings, see \cite{VS94, MS04}. The field equation determining the vortex profile
can be written as
\be
\label{2.0}
f''(r)+\frac{1}{r}f'(r) +(1-\frac{n^2}{r^2})f(r)-f^3(r)=0\,,
\ee
where $f(r)$ is a real function, $f'=df/dr$, and  $n\in {\mathbb Z}$ corresponds to the vorticity.
The physically interesting, globally regular solutions of Eq.\ \eqref{2.0}
satisfy the following boundary conditions (BC):
\be\label{glob-bc}
f(r\to0)=k_nr^n+{\cal O}(r^{n+2})\,,\qquad f(r\to\infty)=f_\infty=1\,,
\ee
and then a major problem of the 2p.\ BVP amounts to find the value(s) of the free parameter,
$k_n$,
to ensure the BC of $f(r)$ at $r=\infty$.

It is not too difficult to integrate numerically
Eq.\ \eqref{2.0} e.g.\ by the ``shooting'' method from $r=0$ to some large value of $r$ and determine
$k_n$ to some precision but it is considerably more difficult to obtain analytical results.
The aim of this Letter is to present
a new analytic procedure to approximate the value of $k_n$ for the connecting trajectory
involving only {\sl algebraic} steps.
Its basic input is the power series expansion of the solution around
a point of analyticity (typically at $r=0$) and the BC at $r=\infty$. With this input
our method reduces the connection problem for $k_n$ to find
the corresponding root of a polynomial equation.
The method is conceptually very simple, it is easy to apply
and moreover it yields good approximations for various ODE's.
Our method is heuristic, we cannot put it on a mathematically rigorous footing as yet,
nor can we precisely define the class of ODE's to which it is applicable.
Nevertheless, it seems to us that with regard
to its simplicity and its large applicability, the new method is of
considerable interest for many applications (it yields with little effort good results for
the connection parameters of many nonlinear ODE's, the
energy levels of the quartic anharmonic oscillator for arbitrary values
of the coupling, etc.).

We illustrate our method in detail on the example of the global vortex \eqref{2.0}.
The first step is to introduce the following sequence of auxiliary first order polynomial (implicit)
ODE's of the form
%
\be\label{algebraic1}
F^{N}(f',f):=f'^N+G_{1}(f)f'^{N-1}+\ldots+G_{N-1}(f)f'+G_{N}(f)=0\,,\ N=1,2\ldots\,,
\ee
where $G_i(f)=\sum_jG_{ij}f^j$ is a polynomial in $f$ with constant coefficients.
Such implicit ODE's
are not easy to handle in general, however, as it will be shown here,
one can squeeze out some important information from the sequence $\{F^{N}(f',f)=0\}$
without having to solve them.
The next step in our method is the determination of the unknown coefficients $G_{ij}$ in $F^{N}(f',f)$.
In order to do this we enforce that Eq.\ \eqref{algebraic1}
be satisfied to the highest possible order in $r$
using the power series
expansion of the solution of Eq.\ \eqref{2.0} $f(r)$, around $r=0$. Having found
the constants $G_{ij}$ this way, we can now impose the BC at $r=\infty$
for $f(r)$ in Eq.\ \eqref{algebraic1} which amounts to
\be\label{algebraic2}
\sum_jG_{Nj}(f_{\scriptscriptstyle\infty})^{\,j}=0\,.
\ee
Since the coefficients $G_{ij}$ depend on $k_n$ (they turn out to be rational functions),
Eq.\ \eqref{algebraic2} represents a polynomial equation for $k_n$.
There is no a priori condition on the degree of the polynomials $G_i(f)$,
we have chosen to impose deg$(G_i)\leq2i$.
We remark that this restriction on the degree is known to be a necessary (but not sufficient) condition for the
absence of movable branch points in Eq.\ \eqref{algebraic1}.
Our main observation is that in the set of real roots, ${\cal S}_N$, of \eqref{algebraic2}
one can find a root, $r_N$, which seems to converge
to the value of the connection parameter and the
corresponding trajectory, $f_N(r)$, of \eqref{algebraic1} yields a global approximation
to the solution of the 2p. BVP of Eq.\ \eqref{2.0}. The main problem in our method is to identify the
``good'' root in ${\cal S}_N$.
At this point we also note that our method performs a kind of resummation from the local
power series expansion through the auxiliary ODE's, nevertheless we see no obvious
relation to more standard resummation techniques such as the Borel technique
or Pad\'e approximants, see the monograph \cite{BO} for a nice review of these and other approximation techniques.

We now show in detail how our method works for the simplest case $N=1$, and for vorticity $n=1$.
Without loosing generality one can assume $k_1>0$.
In order to obtain a nontrivial result for $N=1$ we have to take in Eq.\ \eqref{algebraic1}
for $G_1(f)$ a polynomial of degree two, i.e.\ the simplest auxiliary ODE is a Ricatti eq.:
\be\label{appr1}
f'+G_{10}+G_{11}f+G_{12}f^2=0\,.
\ee
Using the power series expansion of $f(r)$ (by solving Eq.\ \eqref{2.0})
$f(r)=k_1r-k_1 r^3/8+\ldots$,
it is easy to obtain the coefficients $G_{1i}$ in Eq.\ \eqref{appr1}:
$G_{10}=-k_1$, $G_{11}=0$, $G_{12}=3/(8k_1)$. Therefore the solution of Eq.\ \eqref{algebraic2}
corresponding to the BC \eqref{glob-bc} is $k_1=\sqrt{3/8}=0.612\dots$\,,
which constitutes a reasonable first approximation for $k_1$,
(see Table I for the numerical value, $k_{1\rm num}$)
and by solving Eq.\ \eqref{appr1}
one obtains quite a good approximation for the vortex profile function, $f(r)$.

To improve upon this approximation one could keep $N=1$ fixed and increase
only the degree of $G_1(f)$, however, since this scheme is not sufficiently general,
(it works quite well in some, but {\sl not in all} cases), we rather
consider the next member, $N=2$, in our auxiliary ODE sequence with deg$(G_i)=i$.
In this case we have to expand $f(r)$ in Eq.\ \eqref{2.0} up to order $5$, and repeating the
same procedure as for $N=1$, one finds for the coefficients $G_{ij}$:
$G_{10}=k_1(80k_1^2+1)/D_1$, $G_{11}=0$, $G_{20}=-2k_1^2(20k_1^2+7)/D_1$, $G_{21}=0$,
$G_{22}=81/(8D_1)$,
with $D_1=13-40k_1^2$. Therefore Eq.\ \eqref{algebraic2} reduces to
\be\label{N=2}
320k_1^4+112k_1^2-81=0\,,
\ee
whose positive real root is $k_1=\sqrt{(-7+ \sqrt{454})/40}=0.598\dots$,
which compares rather satisfactorily with $k_{1\rm num}$,
considering the simplicity of the calculations.
At this stage it is natural to ask how this approximation
changes keeping $N=2$ fixed but increasing the degree of $G_j$ to $2j$.
Repeating the computations
for this case Eq.\ \eqref{algebraic2} yields
$1 + 368k_1^2 - 2400k_1^4 + 3840k_1^6=0$, which has two real positive roots,
$0.587\ldots$ and $0.531\ldots$. Now  $0.587\ldots$, is a better approximation
than the previous one, however there is also a  second ``spurious'' root, $0.531\ldots$,
which needs to be excluded.
The appearance of such ``spurious'' roots for increasing $N$ and for increasing degrees of
$G_j$ is a general feature and it is a main drawback of the method.
In practice, however, this does not necessarily causes too
serious problems , since one can easily follow the
``good'' roots by continuity
and checking their stability against changing $N$ and the degrees of $G_j$.
In the following, unless indicated otherwise, we shall choose the degree of the polynomials
$G_j$ to be $j$, this fixes the {\sl global degree} of $F^{N}(f',f)$ to be $N$.
In this case there are altogether
$(N+1)(N+2)/2$ constants $G_{ij}$ to be determined, and the order of the power
series in $r$ to be used has to be chosen accordingly.
We present the approximate values of the connection parameter $k_n$ for $n=1,\ldots,4$
up to $N=8$ in Table I.
\begin{table}[ht]
{\caption{Convergence of the approximants for the connection parameters of the global vortex for $n=1,2,3,4$}}
\vskip 0.3cm
\hspace{10mm}%
\begin{tabular}{|c|l|l|l|l|}\hline
$N$ & $k_{1}$ & $k_{  2}$ & $k_{3}$ &$k_{4}$ \\
\hline
                 3  & 0.585    & -                 & -                 & - \\
                 4  & 0.5831   & -                 & 0.021     & - \\
                 5  & 0.58315  & 0.1527            &0.025     & - \\
                 6  & 0.583190  & 0.1529            &0.0264    & 0.0028  \\
                 7  & 0.5831894 & 0.15310           &0.026183 & 0.0034  \\
                 8  & 0.5831894936         & 0.15309           &0.026185 & 0.0033  \\
\hline
     $k_{\rm num}$  & 0.5831894959       & 0.1530991029      & 0.02618342072 &
     0.00332717340 \\
\hline
\end{tabular}
\label{tableau1}
\end{table}
Remarkably for $n=1$ the $N=8$ approximation
yields $8$ correct digits compared with $k_{1\rm num}$.
For increasing values of $n$ the order of the series in $r$ must be also increased for a given
degree of $F^N(f',f)$ and consequently the degree of the polynomial in $k_n$ is greater.
We stress that all computations are analytic, (they have been performed on a standard desktop
computer using Mathematica 5.2), except to find the
roots of the corresponding polynomial where numerical methods became inevitable.

Next we show that our method also yields good results
for an important eigenvalue problem, the determination of the energy levels of the quartic
anharmonic oscillator in $1$ dimension.
Using dimensionless variables the Schr\"odinger equation can be written as:
\be
\label{3.1}
f''(x)+ \left(2 E -\beta x^4-x^2\right) f(x)=0,
\ee
where $\beta$ is the coupling parameter and $E$ is the energy eigenvalue
\footnote{Note that our $\beta$ corresponds to $g/2$ of Ref.\ \cite{kleinert}}.
The 2p.\ BPV
for the ground state wave-function (which is even) can be put in the form $f(x=0)=1$,
$f(x\to\infty)=0$.
For $N=3$ one obtains the following algebraic eq.\ for the eigenvalue $E$
\be
\label{35}
-6408 E^6+12960\beta E^5+2356 E^4-2976 \beta E^3-2(1440\beta^2+133)E^2+168 \beta E+25=0\,.
\ee
The harmonic oscillator corresponds to $\beta=0$, whose exact ground
state energy is $E=1/2$. For $\beta=0$ there is a single positive real root
of Eq.\ \eqref{35}, $0.5166\ldots$, and the choice of the roots for $N=3$, $E_3$, in Table \ref{tableau5}
has been done by following this root as $\beta$ has been varied.
We summarize our results for $N=3\,,4\,,5\ {\rm and}\ 10$ in
Table \ref{tableau5}, where we also compare them with the known ones in
the literature \cite{kleinert} denoted as $E_{\rm v}$ .
\begin{table}[thp]
\begin{tabular}{|c|c|c|c|c||c|}\hline
$\beta$ & $E_3$ &$E_4$ & $E_5$ &  $E_{10}$ & $E_{\rm v}$  \\
\hline
0.2       &0.51   &0.562   &0.5598 &0.5591455  &0.5591463 \\
1         &0.78   &0.704   &0.6984 &0.6961795  &0.6961758  \\
2         &0.90   &0.813   &0.8065 &0.8037773  &0.8037707 \\
4         &1.08   &0.965   &0.9548 &0.9515767  &0.9515685 \\
100       &2.85   &2.56    &2.502  &2.4983125  &2.4997088 \\
400       &4.49   &4.02    &3.931  &3.930989   &3.9309313  \\
2000      &7.65   &6.86    &6.692  &6.694321   &6.6942209  \\
40000     &20.7   &18.6    &18.13  &18.13751   &18.137229  \\
$2\times 10^6$ &76.3& 68.4  & 66.76 & - &-\\
$2\times 10^9$ &763& 684  & 667.6 & - &-\\
\hline
\end{tabular}
\caption{Convergence for the ground state energy of the anharmonic oscillator,
$E_N$ for $N=3\,,4\,,5\ {\rm and}\ 10$}
\label{tableau5}
\end{table}
One can thus see that our method yields very good approximate values for the ground state energy
of the anharmonic oscillator for {\sl all values} of the coupling.
We remark that some of the other real roots of the polynomial in $E$ are related
to the energy levels of the excited states and by our method one can also obtain
approximate values for them, but we will not elaborate on this point here.

Let us next present here some results on the simplest ``skyrmion'' solution, which is of considerable
interest as a good approximation for baryons, we refer to the recent monograph \cite{MS04} for details.
The pertinent ODE for the spherically symmetric skyrmion field can be written as:
\be
(r^2f')'+2f''\sin^2\!f+\sin(2f)\left[f'^2-1-(\sin^2\!f)/r^2 \right]=0\,,
\ee
together with the BC's $f(r)=\pi+kr+{\cal O}(r^3)$, $f\to0$ for $r\to\infty$.
In this case when applying our method we have found that the choice
deg$(G_i(f))=2i$ yields significantly better results than deg$(G_i(f))=i$.
This way we find for the connection parameter: $k = - 2.084$ for $N=2$;
$k = - 1.996$ for $N=3$ and $k = - 2.003$ for $N=4$, whereas $k_{\rm num}=-2.007$.
The agreement is quite satisfactory taking into account
the relatively low degree of the auxiliary equations \eqref{algebraic1}.

An important physical quantity is the total energy of such localized solutions.
For example the energy of the skyrmion in a ball of radius $R$ is given as
$E(R)=\int_0^R{\cal E}dr $, where ${\cal E}$ is the energy density:
\be
{\cal E}=[r^2f'^2+2(1+f'^2)\sin^2\!f+(\sin^4\!f)/r^2]/(3\pi)\,,
\ee
and the {\sl total energy} is then $E(\infty)$.
The direct way to compute approximate values for the energy, i.e.\
finding the corresponding solutions of the auxiliary ODE \eqref{algebraic1} first
and then evaluating $E(R)$ for them
would be quite difficult without resorting to numerics.
We can apply a slight variation
of our method, however, to obtain approximate values for the total energy once
the value of the connection parameter is known.
To do this we consider the sequence $\{F^N({\cal E}, E)=0\}$.
Since the power series expansion of the energy, $E(r)$, is completely fixed for any given
value of the connection parameter, $k$, all the coefficients, $G_{ij}$, will be determined.
Then we solve the corresponding algebraic equation \eqref{algebraic2}
for the unknown value of $E(\infty)$.
This way for $N=9$ we have obtained $E(\infty) = 1.22$, which seems to be quite good
when compared to the numerical value, $E_{\rm num} = 1.23$.

We consider now some ODE's originating from a completely different problem,
the fixed point equation of Wilson's exact renormalisation group (RG).
The RG equation for scalar field theories in the local potential approximation
can be written as \cite{bagn-berv}:
\be
\label{4.1}
2f''(x)-4 f(x)f'(x)-5xf'(x)+f(x)=0,
\ee
where $f(x)= V'(x)-x$, and $V(x)$ is the potential.
The pertinent solution of \eqref{4.1} is an odd function of $x$ and
for $x\to0$ $f(x)=kx+{\cal O}(x^3)$. For large values of $x$
$f(x\to \infty)\to ax^{1/5}$ where $a$ is a constant.
Since in the present case $f_{\infty}=\infty$, $f''\to 0$ and $f'\to 0$ for $x\to\infty$,
it is rather natural to slightly modify the method by considering the auxiliary ODE's
\eqref{algebraic1} for ($f''$,$f'$), i.e.\ $F^N(f'',f')=0$.
Then proceeding exactly as before we obtain a polynomial equation
in $k$, and we find with a good convergence $k=-1.22859876$ for $N=6$.
This agrees quite well with the numerical value $k_{\rm num}=-1.22859820$
\cite{ball-haag-lat-mor}.

Let us consider yet another example, the Wegner-Houghton's fixed point equation
in the local potential approximation \cite{hazen-hazen},
\be
\label{4.3}
2\ln(1+v''(x))+6v(x)-xv'(x)=0\,.
\ee
The change of dependent variable, $f(x)=v'(x)$,
gives the simpler differential equation
\be
\label{4.4}
2f''(x)+[1+f'(x)][5f(x)-x f'(x)]=0\,.
\ee
The solution of interest satisfies the following BC's: $f(x\to 0)=kx+{\cal O}(x^3)$ and
$f(x\to\infty)\to ax^{5}$ where $a$ is a constant.
The connection parameter, $k$, obtained numerically is $k_{\rm num}=-0.461533$ \cite{bagn-berv}.
The problem is similar to the precedent one but we use now higher derivatives,
$F^N(f^{(7)},f^{(6)})=0$.
Proceeding in the same way that in the previous case we obtain
$k=-0.46144...$ for $N=6$ which agrees with the numerical value,
although not so well as in the previous case, probably due
to the use of derivatives of higher order.

We now show that our method can be generalized in a very simple way for a system of $M$ ODE's
for the set of unknowns $\{f_m(r)\}$, $m=1,\ldots, M$.
We introduce
for each unknown function a first order auxiliary implicit ODE $\{F^{N_m}_m(f'_m,f_m)=0\}$,
where $F^{N_m}_m(f'_m,f_m)$ is a polynomial of degree $N_m$ in $f'_m$
(c.f.\ Eq.\ \eqref{algebraic1}).
The constant coefficients in $F^{N_m}_m(f'_m,f_m)$ are determined
by demanding that $\{F^{N_m}_m(f'_m,f_m)=0\}$ be satisfied to the highest possible order in the power
series solutions of $f_m$ at the origin, say. Then proceedings exactly as for the case of a single unknown
we impose the BC at infinity and obtain a system of  algebraic equations
of the form $\sum_jG_{N_mj}(f_m^{\,j}(\infty))=0$ for the connection parameters.
As a concrete illustration we shall consider the field
eqs.\ of the static, rotationally symmetric, gauged vortex
of Nielsen-Olesen \cite{niels-oles} and those of
the 't Hooft-Polyakov magnetic monopole \cite{'tho74}.

The differential equations determining the cylindrically symmetric
magnetic potential resp.\ scalar fields, $a$, $f$
of the Nielsen-Olesen vortex with a single magnetic flux quantum can be written as:
\bsub
\label{5.16}
r(r f')'-f\left[(1-a)^2-r^2\beta(1-f^2)\right]&=&0\,,\\
ra''-a'+2r(1-a)f^2&=&0\,,
\esub
where $\beta$ corresponds to the self-coupling of the scalar field.
The BC's necessary to ensure regularity and finite energy at $r=0$ and at $r=\infty$ are:
$f=f_1=c_{1} r +{\cal O}(r^3)$, $a=f_2=c_{2} r^2+{\cal O}(r^4)$, $f(\infty)=1$, $a(\infty)=1$.
To implement our method we have chosen $F^{N_1}_1(f_1',f_1)$ resp.\
$F^{N_2}_2(f_2',f_2)$
to be polynomials of degree $N_1$ resp.\ $N_2=N_1+1$ in order to have roughly the same number
of terms in the algebraic eqs.\ for $c_1$, $c_2$.
Next we recall the differential equations for the static spherically symmetric magnetic monopoles
in a spontaneously broken SU(2) gauge theory:
\bsub\label{5.1}
(r^2f_1')' - f_1[2f_2^2+\frac{r^2}{2}\beta^2(f_1^2-1)] &=& 0\,,\\
r^2f_2''- f_2[(f_2^2-1)+r^2f_1^2]&=& 0\,,
\esub
where $f_1$ and $f_2$ correspond to the Higgs and the gauge field, respectively.
The regular BC's at $r=0$ resp.\ $r=\infty$ are $f_1=d_1r+{\cal O}(r^3)$, $f_2=1-d_2r^2+{\cal O}(r^4)$,
$f_1(\infty)=1$ and $f_2(\infty)=0$.
We present in Table \ref{tableau9} our results for the connection parameters
of both the vortex and of the magnetic monopole for $N_1=3$ and $N_2=4$, together with
the corresponding numerical values from Refs.\ \cite{FRV, for-ob-re}.
The agreement is good, taking into account that the computations involving
two polynomials are more cumbersome, moreover for both
for small and large values of $\beta$, one needs to increase the values of $N_1$ and $N_2$
more and more.
We recall here that for $\beta=0$ eqs.\ \eqref{5.1} are analytically soluble, yielding $d_1=1/3$, $d_2=1/6$.
\begin{table}[thp]
\begin{tabular}{|c|c|c|c|c|c|c|c|c|}\hline
 $\beta$  &  $c_{1}$ & $c_{2}$  & $c_{1\rm num}$&$c_{2\rm num}$ &$d_1$& $d_2$ &$d_{1\rm num}$& $d_{2\rm num}$\\
\hline
0  &  --& --& --&--&$0.3329$ &   $0.1523$ &$ 1/3$ &  $1/6$  \\
1  &$0.8542 $ &$0.5007$& $0.8532 $  &$0.5000$ & $0.7343$ &$0.3422$ &$ 0.7318$ & $0.3409$  \\
2  &$ 1.0996$ &$0.6169$& $1.0993 $  &$0.6166$ & $1.0692$ &$0.4501$ &$ 1.0683$ & $0.4491$ \\
3  &$ 1.2846$ &$0.6975$& $1.2843 $  &$0.6969$ & $1.4025$ &$0.5350$ &$ 1.4003 $ &$0.5321$ \\
4  &$ 1.4393$ &$0.7609$& $1.4387 $  &$0.7597$ & $1.7406$ &$0.6100$ &$ 1.7339 $ &$0.5997$\\
5  &$ 1.5748$ &$0.8137$& $1.5741 $  &$0.8119$ & $2.0872$ &$0.6901$ &$ 2.0701 $ &$0.6567$\\
6  &$1.6972 $ &$0.8595$& $1.6962 $  &$0.8569$ & $2.4421$ &$0.7834$ &$ 2.4091 $ &$0.7060$ \\
7  &$ 1.8097$ &$0.9001$& $1.8082 $  &$0.8967$ & $2.8035$ &$0.8938$ &$ 2.7503 $ &$0.7492$ \\
10 &$ 2.1052$ &$1.0011$& $2.1024 $  &$0.9944$ & $3.6269$ &$0.9678$ &$ 3.7850 $ & $0.8542$ \\
\hline
\end{tabular}
\caption{ Connection parameters for the vortex and the magnetic monopole with $N_1=3$, $N_2=4$}
\label{tableau9}
\end{table}
We have also computed the total energy for $\beta=1$ applying the same procedure as for the skyrmion,
and we have obtained $E(\infty)= 1.2136$ for $N=10$ with deg$(G_j)=2j$ (c.f.\ $E_{\rm num}= 1.237$ \cite{for-ob-re}).

In conclusion we have presented a new algebraic scheme to obtain the connection parameters,
the energy eigenvalues and the total energy
for a number of physically interesting two point boundary value problems associated with ODE's.
The method is based on a sequence of auxiliary first order implicit polynomial ODE'S,
which is determined from the series expansion of the solution to a given degree.
Imposing the BC's for the solution of the auxiliary ODE's yields
algebraic equations for the connection parameters.
The computation of power series expansions
and finding the roots of the algebraic equations
are easily implementable on symbolic formula manipulation
systems. It seems to be an interesting problem
to clarify the mathematical basis of our method.

{\bf Acknowledgements}
We would like thank Claude Bervillier for pointing out that the fixed
point equations of the RG serve as interesting illustrations of our method
and for his interest.
We also thank him and Peter Breitenlohner
for providing us various numerical results.
One of us (B.B.) had also some interesting discussions with Stam Nicolis.

\end{document}